\documentclass{svproc}
\usepackage{xcolor}
\usepackage{amsmath,amssymb}
\usepackage{hyperref}
\usepackage[left=3.2cm,top=3.2cm,right=3.2cm,bottom=3.2cm]{geometry}
\usepackage{epsfig}
\usepackage{booktabs}
\usepackage{wrapfig}
\usepackage{amssymb}
\usepackage{longtable}
\usepackage{multirow}
\usepackage{array}
\usepackage{paralist}
\usepackage{graphicx}
\usepackage{floatrow}
\usepackage{url}
\usepackage{bm}

\date{}
\begin{document}

\mainmatter

\title{Centrality in dynamic competition networks}

\author{Anthony Bonato\inst{1} \and Nicole Eikmeier\inst{2} \and David F.~Gleich\inst{3}
	\and Rehan Malik\inst{1}}

\institute{Ryerson University, Toronto, Ontario, Canada
\and
Grinnell College, Grinnell, IA, USA
\and
Purdue University, West Lafayette, IN, USA}

\maketitle

\newcommand{\nenote}[1]{\textbf{\textcolor{blue}{\emph{XXX-Nicole-XXX}}} \textcolor{blue!80!white}{#1}
\message{LaTeX Warning: TODO-from-Nicole (\the\inputlineno): #1}}

\begin{abstract}
Competition networks are formed via adversarial interactions between actors. The Dynamic Competition Hypothesis predicts that influential actors in competition networks should have a large number of common out-neighbors with many other nodes. We empirically study this idea as a centrality score and find the measure predictive of importance in several real-world networks including food webs, conflict networks, and voting data from Survivor.
\end{abstract}

\section{Introduction}\label{intro}

While social networks are often studied from the perspective of positive interactions such as friendship or followers, the impact of negative social interaction on their structure and evolution cannot be ignored. Structural balance theory posits positive and negative ties between actors in social networks, and assumes such signed networks will stabilize so that triples of actors are either all mutually friends or possess common adversaries; see \cite{he}, and \cite{k} for a modern treatment. The prediction of the signs of edges in a social network was previously studied~\cite{predict,negative,frenemy}. Further, negative interactions as a model for edges was studied in the context of negatively correlated stocks in market graphs~\cite{market}, and in the spatial location of cities as a model to predict the rise of conflicts and violence~\cite{guo}. Even in the highly cited Zachary Karate club network~\cite{zach}, the negative interaction between the administrator and instructor was the impetus for the split of the club participants into two communities. We propose that competition or negative interactions are critically important to the study of social networks and more broadly, real-world complex networks, and are often hidden drivers of link formation.

In~\cite{dch}, we investigated properties inherent in social networks of competitors that evolve dynamically over time. Such networks are viewed as directed, where a directed edge from nodes $u$ to $v$ corresponds to some kind of negative social interaction. For example, a directed edge may represent a vote by one player for another in a social game such as the television program Survivor. Directed edges are added over discrete time-steps in what we call dynamic competition networks. Our main contribution in~\cite{dch} was the presentation of a hypothesis, referred to as the Dynamic Competition Hypothesis, or (DCH), that served as a predictive tool to uncover alliances and leaders within dynamic competition networks. We provided evidence for the hypothesis using U.S.\ voting record data from 35 seasons of Survivor.

In the present paper, we focus on a particular implication of the DCH. Namely, the DCH predicts that leaders and central actors in these networks should have large a large number of common out-neighbors with other nodes in the network. Consequently, this score should constitute a more accurate and interesting centrality score in competition networks where edges have a negative connotation. We study this score in terms of its ranking of leaders in various kinds of networks ranging from additional international seasons of Survivor, to conflict networks, and to food webs.


We organize the discussion in this paper as follows. In Section~\ref{DCH}, we formally define dynamic competition networks, and review the DCH as stated in~\cite{dch}, with a focus on the common out-neighbor scores, called CON scores. In Section~\ref{data}, we investigate using CON scores as centrality measures in three distinct sources: i) voting data from all international (that is, non-U.S.) seasons of Survivor, ii) from conflict networks arising from the countries of Afghanistan, India, and Pakistan, and iii) in 14 food webs. We find that the CON scores predict influential actors in the networks with high precision.
The final section interprets our results for real-world complex networks, and suggests further directions.

We consider directed graphs with multiple directed edges throughout the paper. Additional background on graph theory and complex networks may be found in the book~\cite{bbook} or \cite{bt}.

\section{The Dynamic Competition Hypothesis }\label{DCH}

The Dynamic Competition Hypothesis (DCH) provides a quantitative framework for the structure of dynamic competition networks. We recall the statement of the DCH as first stated in \cite{dch}. Before we state the DCH, we present some terminology.

A \emph{competition network} $G$ is one where nodes represent actors, and there is directed edge between nodes $u$ and $v$ in $G$ if actor $u$ is in competition with actor $v$. A \emph{dynamic competition network} is a competition network where directed edges are added over discrete time-steps. For example, nodes may consist of individuals and edges correspond to conflicts between them; as another example, we may consider species in an ecological community, and directed edges correspond to predation. Observe that dynamic competition networks may have multiple edges if there were multiple conflicts; further, not all edges need be present.

The central piece of the DCH we study here are the \emph{common out-neighbor} scores. Without loss of generality, we assume that the node correspond to integers such that we can use the nodes to address an adjacency matrix as well. Consequently, let $\bm{A}$ be the adjacency matrix of given competition network. Entries in the matrix are $0$ or positive integers for the number of competition interactions. For nodes $u,$ $v,$ and $w,$ we say that $w$ is a \emph{common out-neighbor} of $u$ and $v$ if $(u,w)$ and $(v,w)$ are directed edges. Alternatively, $A_{uv} A_{vw} \ge 1$. For a pair of distinct nodes $u,v$, we define $\mathrm{CON}(u,v)$ to be the number of common out-neighbors of $u$ and $v$. Note that this common out-neighbor score counts multiplicities based on the minimum number of interactions:  $\mathrm{CON}(u,v) = \sum_{k} \text{min}(A_{uk}, A_{vk})$, which corresponds to a multiset intersection.
For a fixed node $u$, define $$\mathrm{CON}(u)=\sum_{v\in V(G)} \mathrm{CON}(u,v).$$ We call $\mathrm{CON}(u)$ the \emph{CON score} of $u.$  For a set of nodes $S$ with at least two nodes, we define $$\mathrm{CON}(S) = \sum_{u,v\in S} \mathrm{CON}(u,v).$$ Observe that $\mathrm{CON}(S)$ is a non-negative integer.

In the DCH,  \emph{leaders} are defined as members of a competition network with high standing in the network, and edges emanating from leaders may influence edge creation in other actors. In the context of conflict networks within a country, leaders may be actors who exert the strongest political influence within the country; note that these may not be the largest or most powerful actors. As another example, leaders in a food web would naturally have higher trophic levels (that is, higher position in a food chain). The DCH characterizes leaders as those nodes with high CON scores, low in-degree, high out-degree and high closeness. Recall that for a strongly connected digraph $G$ and a node $v$, we define the \emph{closeness} of $u$ by $$C(u)=\left(\sum _{v \in V(G)\setminus \{u \}}  d(u,v)\right)^{-1}$$  where $d(u,v)$ corresponds to the distance measured by one-way, directed paths from $u$ to $v$.

In this paper, we focus on the implication that leaders in competition networks should have high CON scores, which suggests this is a natural centrality measure for these networks. The DCH also involves the notion of alliances, that does not factor into our present study.  \emph{Alliances} are defined as groups of agents who pool capital towards mutual goals. In the context of social game shows such as Survivor, alliances are groups of players who work together to vote off players outside the alliance. Members of an alliance are typically less likely to vote for each other, and this is the case in strong alliances. This is characterized in terms of \emph{near independent sets}; see~\cite{dch} for the formalism.

In summary, the \emph{Dynamic Competition Hypothesis} (or \emph{DCH}) asserts that dynamic competition networks satisfy the following four properties.
\begin{enumerate}
	\item Alliances are near independent sets.
	\item Strong alliances have low edge density.
	\item Members of an alliance with high CON scores are more likely leaders.
	\item Leaders exhibit high closeness, high CON scores, low in-degree, and high out-degree.
\end{enumerate}

Our focus in this work will be on the validation of the DCH with regards to detecting leaders; in particular, we will focus on items (3) and (4) of the DCH. Note that while we expect leaders to be in alliances (that is, have prominent local influence), leaders are determined via global metrics of the network.

\section{Methods and Data}\label{data}

\subsection{Survivor}
In \cite{dch}, we studied the voting history of U.S.\ seasons of Survivor, which is a social game show where players compete by voting each other out. In Survivor, strangers called survivors are placed in a location and forced to provide shelter and food for themselves, with limited support from the outside world. Survivors are split into two or more tribes which cohabit and work together. Tribes compete for immunity and the losing tribe goes to tribal council where one of their members is voted off. At some point during the season, tribes merge and the remaining survivors compete for individual immunity. Survivors
voted off may be part of the jury. When there are a small number of remaining survivors who are finalists (typically two or three), the jury votes in favor of one of
them to become the Sole Survivor who receives a cash prize of one million dollars. Figure~1 represents a graphical depiction of the voting history of a season of U.S.\ Survivor.

\begin{figure}[h!]
\begin{center}
\epsfig{figure=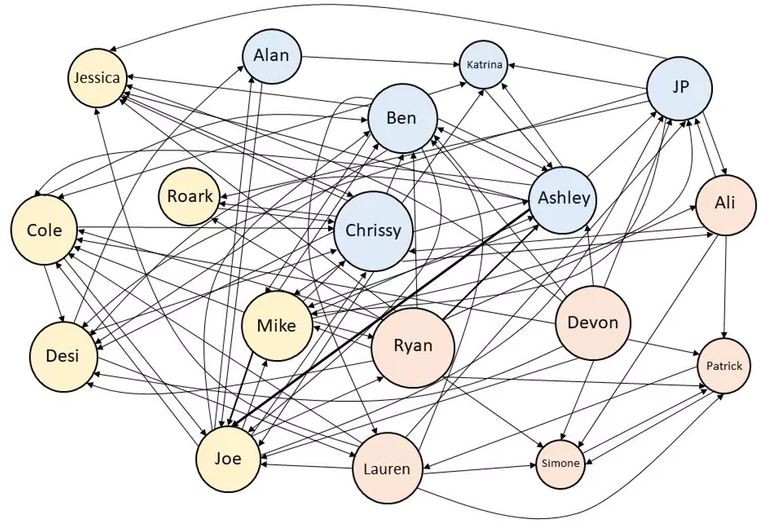,scale=0.5}
\caption{The Survivor Heroes vs. Healers vs. Hustlers co-voting network. Nodes are scaled by closeness, and color-coded according to their original tribe. Thicker edges represent multiple votes.}
\end{center}
\end{figure}

\begin{table}[h!]
\raisebox{-\height}{\begin{tabular}{ l  l  l  l  c }
\multicolumn{5}{c}{Australian Survivor (2002)} \\
Name & ID & OD & C & CON \\ \toprule
Robert&5&10&0.714&44\\
Sciona&1&9&0.652&37\\ \midrule
Joel&7&8&0.625&35\\
Katie&3&9&0.652&38\\
Sophie&3&8&0.652&38\\
Jane&9&6&0.625&36\\
Lance&8&5&0.577&27\\
Craig&8&8&0.577&18\\
Naomi&8&7&0.5&25\\
Caren&10&6&0.5&25\\
Sylvan&3&5&0.417&30\\
Deborah&4&4&0.395&23\\
Jeff&5&1&0.395&4\\
David&6&3&0.441&23\\
Tim&4&2&0.294&10\\
Lucinda&8&1&0.0&7
\end{tabular} }
\raisebox{-\height}{\begin{tabular}{ l  l  l  l  c }
\multicolumn{5}{c}{Robinson 2009} \\
Name & ID & OD & C & CON \\\toprule
Ellenor&0&6&0.563&36\\
Jarmo&7&8&0.557&34\\ \hline
Anna&2&10&0.645&54\\
Nina&4&7&0.557&38\\
Erik Bl.&7&9&0.612&47\\
Lukas&4&7&0.557&31\\
Angela&6&8&0.612&46\\
Ranjit&5&5&0.51&30\\
Christian&3&4&0.51&24\\
Rafael&5&4&0.49&28\\
Erik Bi.&9&5&0.438&26\\
Erik R.&5&4&0.422&18\\
Mika&5&3&0.306&13\\
Josefine&0&2&0.265&17\\
Erika&7&2&0.306&15\\
Beatrice&6&2&0.35&12\\
Micha&12&1&0.299&7\\
\end{tabular} }
\raisebox{-\height}{\begin{tabular}{ l  l  l  l  c }
\multicolumn{5}{c}{Survivor South Africa Malaysia} \\
Name & ID & OD & C & CON \\\toprule
Lorette&4&9&0.653&35\\
Grant&5&8&0.653&33\\ \toprule
Amanda&4&8&0.622&26\\
Mandla&0&8&0.652&32\\
Angie&6&9&0.688&31\\
Angela&4&6&0.594&28\\
Dyke&4&6&0.568&17\\
Hein&5&4&0.484&22\\
Irshaad&3&5&0.544&25\\
Lisa&11&4&0.484&16\\
Rijesh&4&3&0.408&13\\
Nichal&6&2&0.363&12\\
Elsie&8&2&0.436&9\\
Viwe&5&2&0.344&11\\
Nicola&5&1&0.304&6\\
Nomfundo&4&1&0.335&8\\
\end{tabular} }
	\caption{Three international Survivor seasons. Players are listed by first name, in order from top to bottom with the winner at the top, and the first eliminated player at the bottom. For each player we list the in-degree, out-degree, closeness, and CON score. The horizontal line separates finalists from the rest of the group.}
	\label{tab:ExampleSeasons}
\end{table}

We extend the analysis of the 35 U.S.\ seasons in \cite{dch} to 82 international seasons of Survivor. Data used in our analysis was obtained from the Survivor wiki pages \url{https://survivor.fandom.com/wiki/Main_Page}. Several seasons (beyond the 82) were excluded for varying reasons. In some cases, a wiki page exists, but there was no voting data. In other cases, much of the voting information was missing, or the rules are significantly different than the traditional version of the game shows. Nevertheless, the number of seasons collected exceeds the number in \cite{dch}.

In Table~\ref{tab:ExampleSeasons} we display some of the CON scores for a few example seasons. We distinguish which players are finalists, since the rules change in determining who is the last player eliminated. For example, instead of eliminating the last player via votes \emph{against} players, in survivor many players may return for a final vote \emph{for} who they would like to win.

\begin{table}[h!]
	\begin{center}
		\begin{tabular}[H]{ m{2.2cm}  m{1.4cm} m{1.4cm}  m{1.8cm}  m{1.9cm}  m{2cm} }
			& & \textbf{CON} & \textbf{PageRank} & \textbf{Jaccard Similarity} & \textbf{random set} \\ \hline
			\textbf{Survivor} & {\small\textbf{Top 3}}  & \textbf{57.3}     & 43.9  & 47.6 & 11.1-27.3 \\
			&{\small\textbf{Top 5}}& \textbf{81.7}    & 78.0 & 72.0 & 18.5-45.5 \\
		\end{tabular}
	\end{center}
	\caption{Statistics on international Survivor seasons.}
	\label{tab:SBB}
\end{table}

In Table~\ref{tab:SBB}, we detail relevant statistics on these networks. For each network, we consider whether the winner of the season had one of the top three or top five CON scores and list the percentage of such networks. For example from Table~\ref{tab:SBB} we see that 81.7 percent of Survivor winners  had one of the largest 5 CON scores. For comparison, we also compute PageRank (on the \emph{reversed-edge} network, where we change the orientation of directed edge) and Jaccard similarity scores, which are both standard ranking scores. Jaccard similarity is a type of normalized CON score; see \cite{gower}. We find that the CON scores are a more accurate predictor for determining finalists of Survivor than both PageRank and Jaccard similarity. We observe that these results are consistent with the analysis in \cite{dch} for the U.S.\ seasons. As an added comparison, we list the probability of the winner appearing in a random set of three or five players; note that there is a range of percentages depending on how many players are in a given season. In the interest of space, we refer the reader to \url{https://eikmeier.sites.grinnell.edu/uncategorized/competition-show-data/} where we house all data on these seasons.

\subsection{Political Conflicts}
For our second competition network, we extracted data from \emph{The Armed Conflict Location and Event Data Project} (or \emph{ACLED}), which may be found at \url{https://www.acleddata.com/data/}. ACLED catalogs information about political conflict and protests across the world. In these \emph{conflict networks}, nodes correspond to actors in a given region, and edges correspond to conflicts between the actors. Many types of metadata are recorded corresponding to each event. Our particular interest is in the actors involved in each event, and where the event took place. More information about this project can be found on the project website.

An important note about the ACLED data is that we do not know which actor \emph{initiated} a given event. Therefore, we do not consider the majority of edges (events) to be directed. The only events which we assume knowledge about directed-ness is when civilians are involved. 
We restricted our study to a set of events to a particular country; keeping the scale at the country level allows us to keep a larger set of actors. We selected three countries that have a large number of actors and events to analyze: Afghanistan, India, and Pakistan.

We first consider the rankings for Pakistan with commentary.

\begin{figure}[tbh]
	\begin{center}
		\includegraphics[width=\linewidth,clip=true,trim= 0 6.0cm 0 0 ]{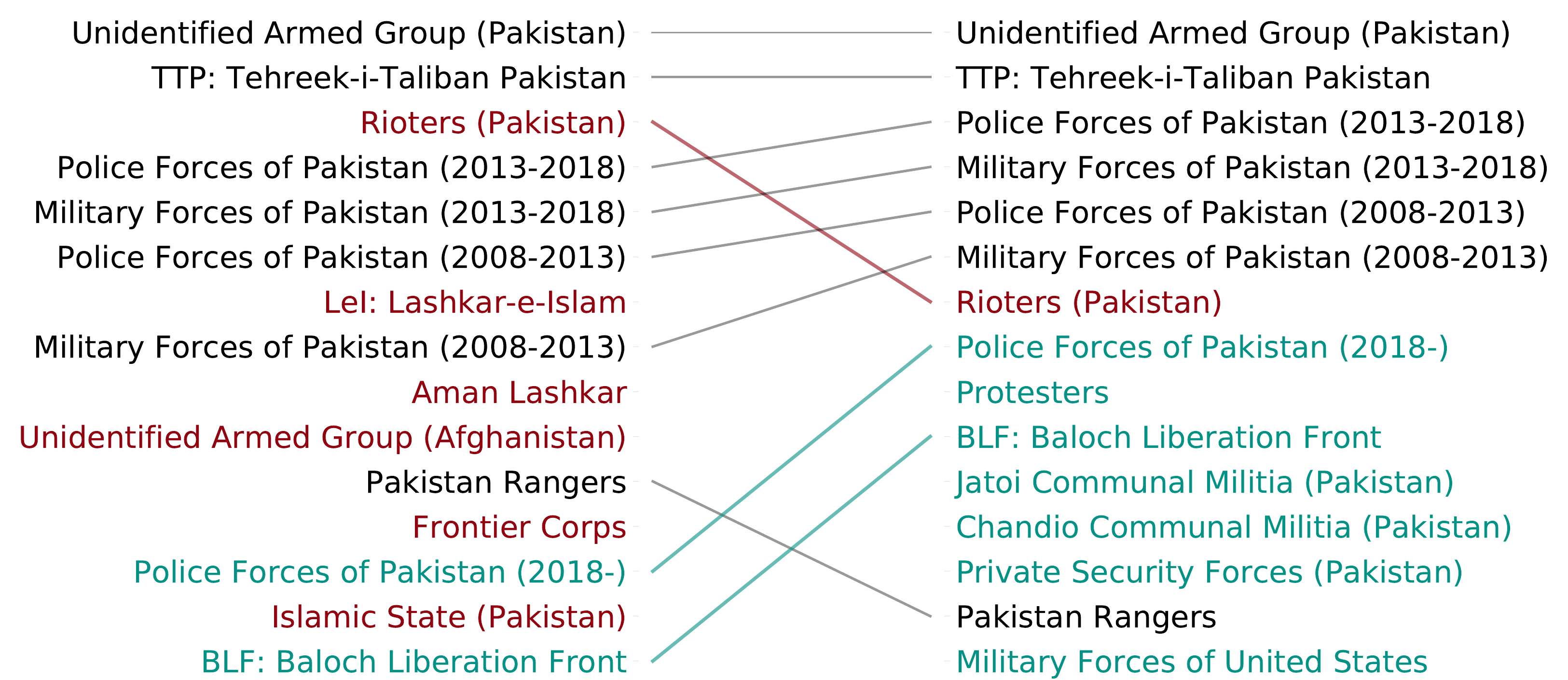}
		\caption{A Slope Graph to compare the rankings via CON and PageRank. On the left, the top actors in the conflict network Pakistan via CON metrics, while on the right, the top actors in Pakistan via PageRank on the reverse network. Actors are labeled in black if the difference in rankings is less than or equal to three. Actors are labeled in red if the CON ranking is at least four places higher than the PageRank, and in green if the PageRank is at least 4 places higher than the CON ranking. Note that no line appears to connect the left and right side if the actor does not show up in the top 15 of the other ranking.}
		\label{fig:Pakistan}
	\end{center}
\end{figure}

Pakistan has faced terrorism activities since 2000, with many militant groups attacking civilians and Pakistan armed forces. TTP (Pakistan) is one of the largest radical extremist groups, which is an umbrella organization of many militant groups such as Lashkar-e-Islam, Islamic State (Pakistan), and Jamaat-ul-Ahrar. In Figure~\ref{fig:Pakistan}, we find that TTP has one of the highest CON scores. TTP has alliances with another terrorist organizations in Pakistan and neighboring countries, which lends to its prominence. In addition, due to the Afghan war, TTP has a strong influence and hold over many Islamic institutions in Pakistan. The Police Forces of Pakistan and Military Forces of Pakistan ensure national security, and they share information for achieving their goals. The Police Forces of Pakistan are an influential actor in the conflict network with another one of the highest CON scores. They perform their duties in all provinces of Pakistan with the help of their paramilitary forces such as Pakistan Rangers and Frontier Corps, and they maintain law and order, as well as border control. Military Forces of Pakistan (2013-2018) has one of the largest CON scores owing to their increased activities against terrorist groups in recent years.

We also offer commentary on some of the lower ranked actors. The Baloch Liberation Front (or BLF) is an ethnic-separatist political front and militant organization that is currently fighting against the Pakistani government for an independent Balochi state. The BLF is the strongest and most influential militant group of Baluchistan, but there has been no confirmed coordination between the BLF and other Balochi and non-Balochi groups, and they operate independently of one another. This is a large reason that BLF have low CON and closeness scores. The Islamic State is a part of the militant Islamist group: Islamic State of Iraq and Levant (ISIS). The Islamic State was formed by some of the TTP leaders and is more successful in Afghanistan. This organization has had less success in Pakistan largely carrying out isolated, small scale attacks. The Police Forces of Pakistan actively participated with the support of paramilitary forces of Pakistan in 2008-2018 for war against terror.  The Police Forces of Pakistan mostly work to maintain the daily law and order in their respective provinces. Likely for these reasons, they have lower CON scores than the years between 2008-2018.

\begin{figure}[tbh]
	\begin{center}
	\includegraphics[width=\linewidth,clip=true,trim= 0 3.0cm 0 0 ]{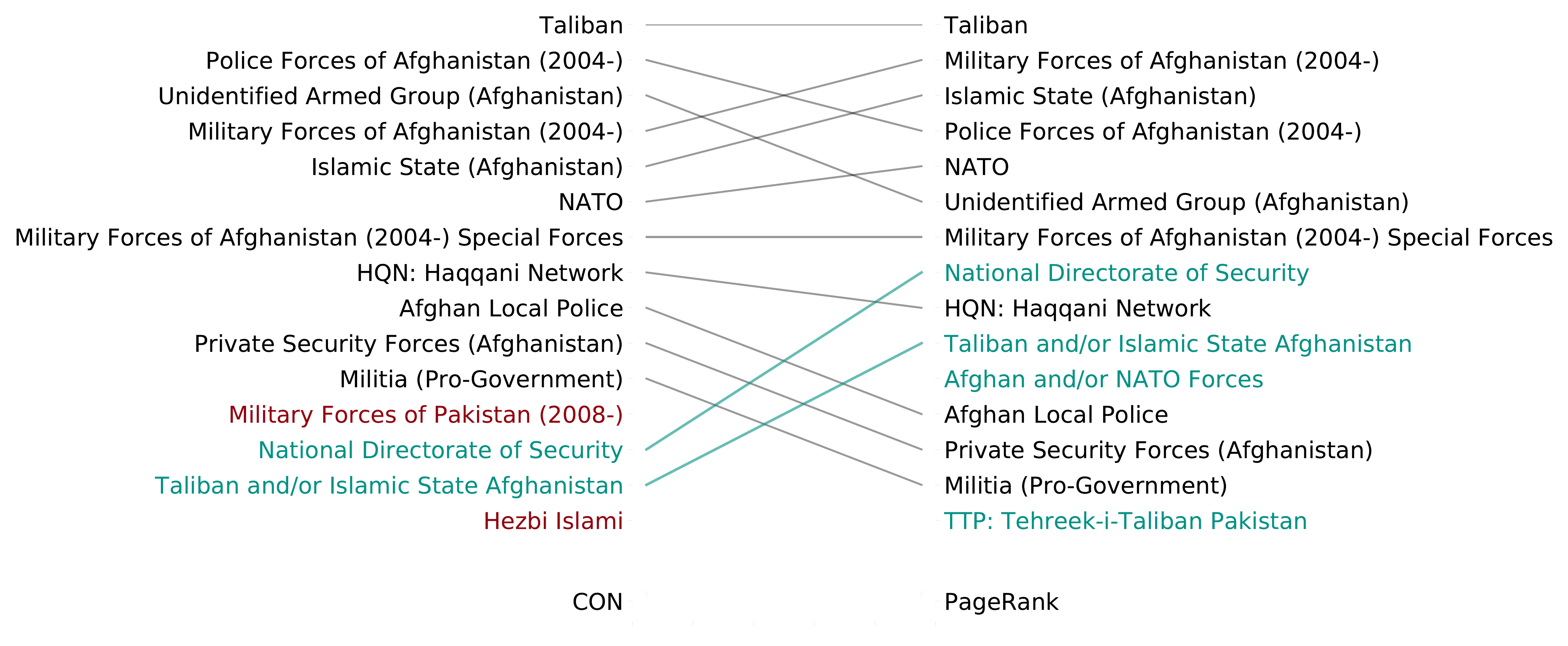}
	\caption{Top actors in Afghanistan via CON score and PageRank.}
			\label{fig:Afgh}
	\end{center}
\end{figure}

\begin{figure}[tbh]
	\begin{center}
	\includegraphics[width=\linewidth,clip=true,trim=0 5.5cm 0 0 ]{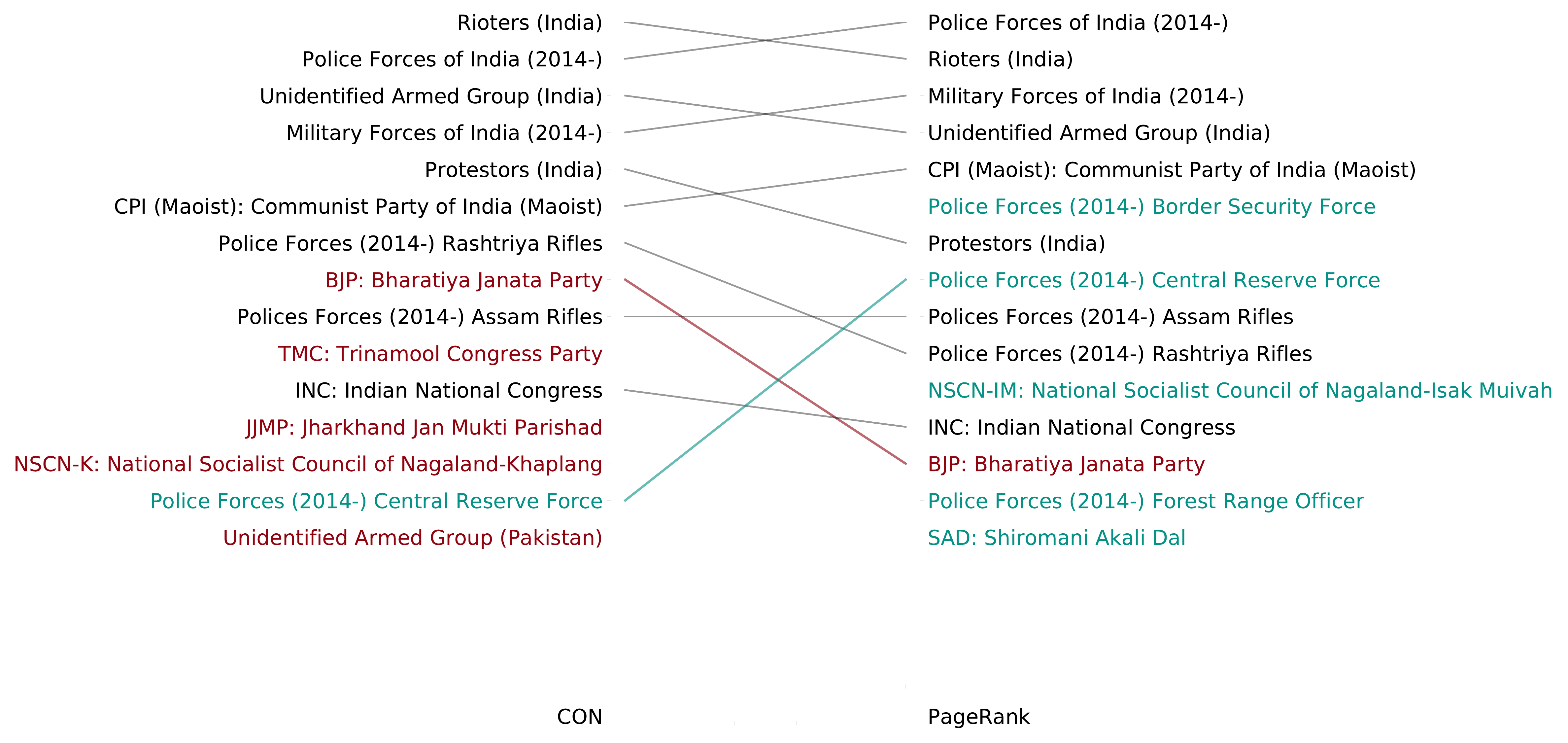}
	\caption{Top actors in India via CON score and PageRank.}
			\label{fig:India}
	\end{center}
\end{figure}

We note that the ranking of the top actors using the CON score (on the left in Figure~\ref{fig:Pakistan}) is not dissimilar to the one using PageRank on the reversed-edge network (on the right in Figure~\ref{fig:Pakistan}). To quantify the difference in the rankings we used Spearman's rank correlation coefficient. Note that we cannot use Pearson correlation because our data is not at all Gaussian. Recall that Spearman's correlation coefficient is defined as
\[ 1 - \frac{6 \sum_{i = 1}^{N} d_i^2}{N(N^2-1)},  \]
where $N$ is the total number of actors, and $d_i$ is the difference in rankings between actor $i$. A value close to 1 means that the two rankings are very well positively correlated. The Spearman correlation for Pakistan is -0.341, which suggests that the rankings are not that similar. In fact, the negative value implies that as the CON ranking decreases, the PageRank score increases. There are 741 total actors we consider in the Pakistan data set, and the later rankings clearly vary greatly.

We finish this section with the rankings for Afghanistan and India in Figures~\ref{fig:Afgh} and~\ref{fig:India}. The Spearman coefficients are 0.604 and -.267 respectively, indicating that the rankings provided by CON matches more similarly to PageRank in the Afghanistan dataset. While we do not provide in-depth commentary on these rankings, we find influential actors in both countries with the largest rankings against DCH metrics.

\subsection{Food Webs}

As a third and final type of data that we analyzed against the DCH, we studied food web datasets from the Pajek website: \url{http://vlado.fmf.uni-lj.si/pub/networks/data/bio/foodweb/foodweb.htm}~\cite{pajek}. These are 14 food webs in total. In food webs, the nodes are species, and a weighted edge $(u,v)$ exists with weight $w$ if $u$ inherits carbon from $v$ (that is, $u$ preys on $v$)~\cite{baird1989seasonal}. We interpret this as a directed negative interaction from node $u$ to node $v$. A noteworthy difference in these networks (vs. Survivor, say) is that  the movement of energy is \emph{balanced}, meaning the in-degree and out-degree for each species is equal.

\begin{figure}[h!]
	\begin{center}
	\includegraphics[width=4.5in,clip=true,trim=0 1.4cm 0 0 ]{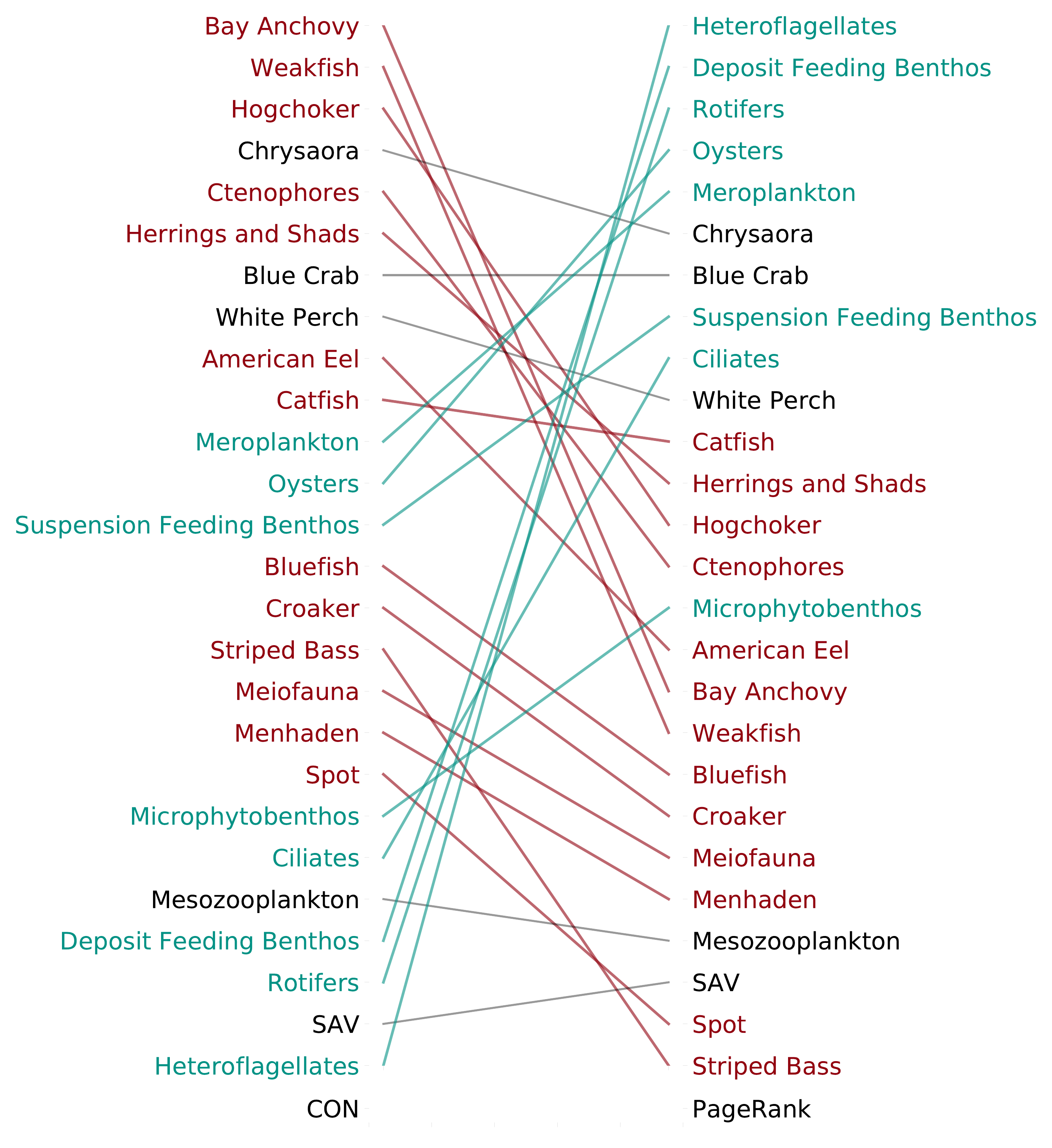}
	\caption{The Chesapeake Bay Lower food web dataset. On the left, organisms are listed in decreasing order, with the largest CON score at the top. On the right, organisms are listed by decreasing PageRank score.}
			\label{fig:ches}
	\end{center}
\end{figure}

\begin{figure}[h!]
	\begin{center}
	\includegraphics[width=4in,clip=true,trim= 0 1.1cm 0 0cm]{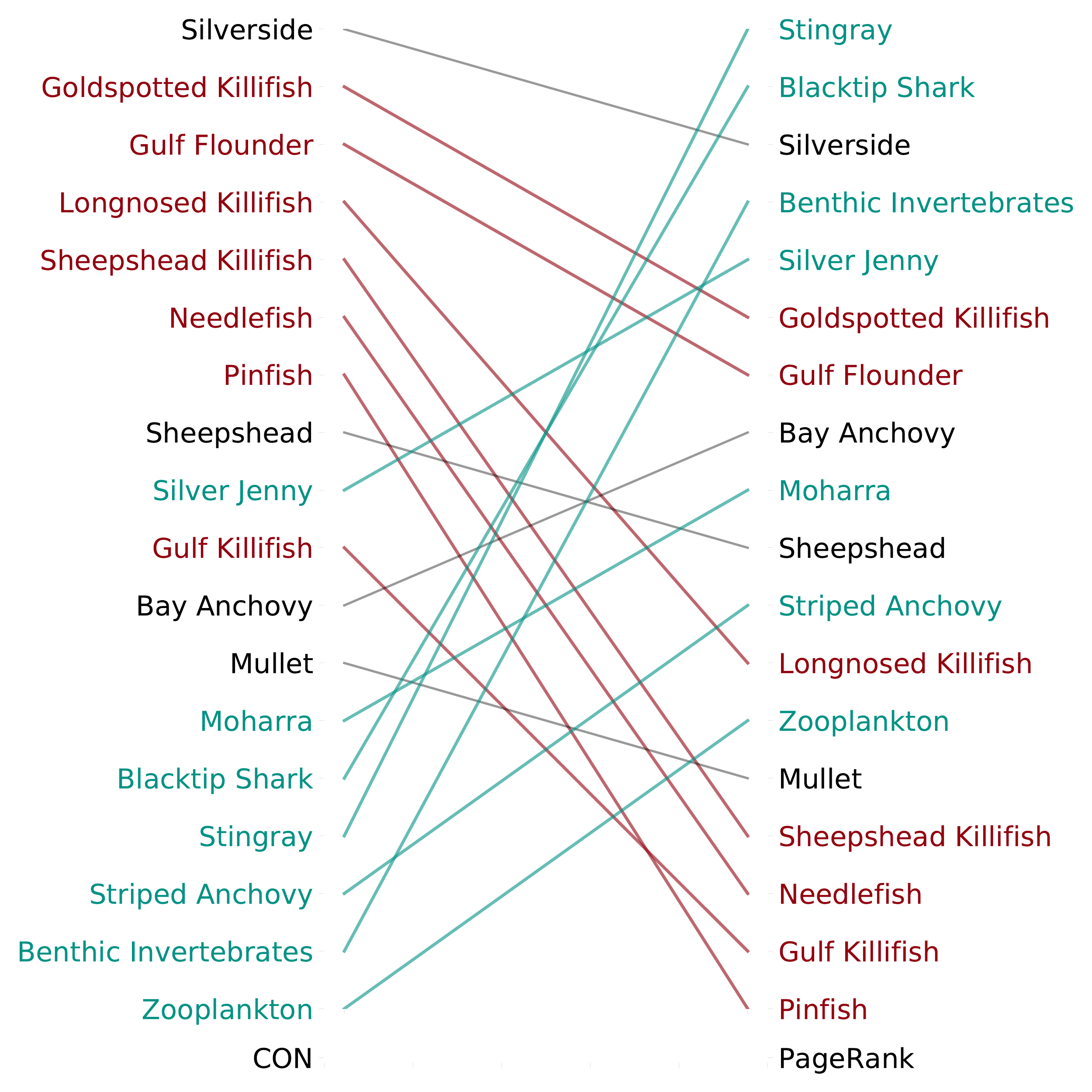}
	\caption{The CrystalC food web dataset. On the left, organisms are listed in decreasing order, with the largest CON score at the top. On the right, organisms are listed by decreasing PageRank score.}
		\label{fig:crystal}
	\end{center}
\end{figure}

Rankings of selected food web datasets are in Figures~\ref{fig:ches} and~\ref{fig:crystal}; the rankings for all the datasets may be found at \url{https://eikmeier.sites.grinnell.edu/uncategorized/competition-show-data/} along with the computed CON scores, closeness, and PageRank on the reversed-edge network.

In studying the rankings of these 14 food webs, we see a difference between the CON rankings and PageRank. PageRank has been used to study the importance of species in regards to co-extinction~\cite{allesina2009-PRFoodwebs,stouffer2011-compartmentalization,McDonald-FoodSurvey}, which we expect is likely reflected in the rankings we see here using vanilla PageRank. However, we find a substantially different ranking when using the CON scores; for example, see the placement of Heteroflagellates in Figure~\ref{fig:ches}. The average Spearman correlation coefficient across these 14 datasets is 0.271, and the range is between 0.004 and 0.554. (Recall that a value close to 1 means very well correlated.) Therefore, we suggest that the CON scores are giving a \emph{different} ranking, which is much closer to \emph{trophic} levels of species. In particular, the CON scores reflect a natural hierarchical structure in ecosystems, and this is consistent with the DCH.

\section{Conclusion and future directions}

We studied an implication of the Dynamic Competition Hypothesis (DCH) for competition networks across several different types of real-world networks. We found that the DCH prediction that high CON scores should correspond to leaders is supported  in predicting winners in international seasons of Survivor, in predicting species with high trophic level species in food web, and for determining influential actors in conflict networks in Afghanistan, India, and Pakistan. Metrics such as CON scores outperformed PageRank as an indicator of influential actors in the competition networks we studied.

While our results provide support for the DCH, more work needs to be done. We did not address items (1) and (2) of the DCH regarding alliances in our data sets, and that would be an important next step. Another direction is to consider an aggregate score, based on the CON score, closeness, and in- and out-degree, as a measure of detecting leaders in competition networks. An interesting direction would be to study more closely the dynamic aspects of competition networks, analyzing them over time to predict leaders. For example, we could analyze the co-voting network of Survivor of each episode of a season, and determine if temporal trends in network statistics predict finalists.

An open question is whether CON score centrality is applicable to large-scale networks exhibiting adversarial interactions, such as in Epinions and Slashdot (which give rise to signed data sets with tens of thousands of nodes, and available from~\cite{lesk}). Epinions was an on-line consumer review site, where users could trust or distrust each other. Slashdot is a social network that contains friend and foe links. A challenge with these data sets from the view of validating the DCH is that there is no inherently defined ranking, as there is in Survivor (via the order contestants were voted off), food webs (trophic level), and in conflict graphs (via political and strategic influence).

\section{Acknowledgments} The research for this paper was supported by grants from NSERC and Ryerson University. Gleich and Eikmeier acknowledge the support of NSF Awards IIS-1546488, CCF-1909528, the NSF Center for Science of Information STC, CCF-0939370, and the Sloan Foundation.

\end{document}